\documentclass[aps,nofootinbib,tightenlines]{revtex4}
\usepackage{stmaryrd}
\usepackage{epsfig}
\usepackage{amsfonts}
\usepackage{amsmath}
\usepackage{graphicx}
\usepackage{color}
\usepackage{amssymb,amsmath,psfrag,slashed,graphicx}

\allowdisplaybreaks[4]

\def \t {\tilde}

\newif\ifContLineOne
\newif\ifContLineTwo
\newif\ifContLineThree

\def\conC#1{\vbox{\ialign{##\crcr
  \ifContLineThree\hrulefill\else\vphantom{\hrulefill}\fi\crcr
  \noalign{\kern3.2pt\nointerlineskip}
  \ifContLineTwo\hrulefill\else\vphantom{\hrulefill}\fi\crcr
  \noalign{\kern3.2pt\nointerlineskip}
  \ifContLineOne\hrulefill\else\vphantom{\hrulefill}\fi\crcr
  \noalign{\nointerlineskip}
  $\hfil\textstyle{\vbox to 14pt{}#1}\hfil$\crcr}}}

\def\DrawLeg#1#2{
  \kern-.2pt              
  \dimen2 =#1             
  \advance\dimen2 by 2pt  
  \dimen3 = 10.6pt        
  \dimen4 =3.6pt          
  \advance\dimen3 by -\dimen2
  \multiply\dimen4 by #2
  \advance\dimen3 by \dimen4
  \raise\dimen2 \hbox{\vrule height\dimen3 width .4pt} 
  \kern-.2pt}             

\def\begC#1#2{\setbox0 =\hbox{$\textstyle{#2}$}
  \dimen0=.5\wd0 \dimen1=\ht0
  \conC{\hskip\dimen0}
  \count255=#1
  \ifnum\count255 =1 \ContLineOnetrue\else
  \ifnum\count255 =2 \ContLineTwotrue\else
  \ifnum\count255 =3 \ContLineThreetrue\fi\fi\fi
  \DrawLeg{\dimen1}{\count255}
  \conC{\hskip\dimen0}
  \kern-\dimen0\kern-\dimen0 \box0}

\def\endC#1#2{\setbox0 =\hbox{$\textstyle{#2}$}
  \dimen0=.5\wd0 \dimen1=\ht0
  \conC{\hskip\dimen0}
  \count255=#1
  \ifnum\count255 =1 \ContLineOnefalse\else
  \ifnum\count255 =2 \ContLineTwofalse\else
  \ifnum\count255 =3 \ContLineThreefalse\fi\fi\fi
  \DrawLeg{\dimen1}{\count255}
  \conC{\hskip\dimen0}
  \kern-\dimen0\kern-\dimen0 \box0}

\begin{document}

\title{On the power divergence in quasi gluon distribution function}

\author{Wei Wang\footnote{wei.wang@sjtu.edu.cn} and Shuai Zhao\footnote{shuai.zhao@sjtu.edu.cn} }
\affiliation{  INPAC, Shanghai Key Laboratory for Particle Physics
and Cosmology,\\
MOE Key Laboratory for Particle Physics, Astrophysics and Cosmology,\\
 School of Physics and Astronomy, Shanghai Jiao Tong
University, Shanghai, 200240,   China}

\begin{abstract}

Recent perturbative calculation of quasi gluon distribution function at
one-loop level shows  the existence of extra  linear ultraviolet divergences in the cut-off scheme.   We employ the auxiliary field approach, and study the renormalization of  gluon operators.  The  non-local gluon operator can mix with new operators under renormalization, and the linear
divergences in quasi distribution function can be into the newly introduced operators. After including the mixing, we find the improved quasi gluon distribution functions  contain only logarithmic
divergences, and thus can be used to extract the gluon distribution in large momentum effective theory.

\end{abstract}

\maketitle

\section{Introduction}

The  high energy behaviour of  a hadron is encoded into  parton
distribution functions (PDFs). Because of the
non-perturbative nature, the PDFs can not be
calculated in QCD perturbation theory. Traditionally, PDFs are extracted using the   experimental data
for the processes in which the factorization theorem
is established~\cite{Butterworth:2015oua,Hou:2016nqm,Hou:2017khm,Gao:2017yyd}.
Another  popular non-perturbative
method is to use lattice QCD (LQCD). However, since   PDFs are defined as light-cone correlation matrix elements~\cite{Collins:2011zzd}, it is a formidable task to calculate the PDFs on the lattice. Instead only the lowest few moments are explored due to the technical difficulties.

A novel approach, named as large momentum effective theory (LaMET), has
been proposed to extract light-cone observables from lattice
calculation~\cite{Ji:2013dva,Ji:2014gla}. In this approach, one can calculate the corresponding quasi observables on the
lattice,  defined by equal-time correlation matrix
elements. The quasi   and light-cone observables share the same infrared (IR) structures. The difference between them is   IR independent,  and   can be calculated perturbatively. Factorization of the  quasi quark distribution has been examined at one-loop level  and proved to
hold to all orders \cite{Xiong:2013bka,Ma:2014jla}.
Therefore, if the matching
coefficient is calculated in perturbation theory, and the quasi
distribution function is evaluated on the lattice, one can extract
light-cone distribution functions. This approach has been
studied extensively
\cite{Gamberg:2014zwa,Ji:2014hxa,Jia:2015pxx,Ji:2015qla,Ji:2015jwa,Monahan:2015lha,Xiong:2015nua,Gamberg:2015opc,Ishikawa:2016znu,Monahan:2016bvm,Bacchetta:2016zjm,
Ji:2017rah,Nam:2017gzm,
Rossi:2017muf,Stewart:2017tvs,Xiong:2017jtn,Hobbs:2017xtq,Broniowski:2017wbr,Carlson:2017gpk,Chen:2017mie,Monahan:2017hpu}, and some encouraging progresses on LQCD calculation are
reported~\cite{Lin:2014zya,Alexandrou:2015rja,Chen:2016utp,Alexandrou:2017huk,Lin:2017ani,Briceno:2017cpo,Chen:2017mzz,Constantinou:2017sej,Zhang:2017bzy}. Some other frameworks, e.g.,
the pseudo-PDFs~\cite{Radyushkin:2016hsy,Radyushkin:2017cyf,Radyushkin:2017ffo,Radyushkin:2017gjd,Orginos:2017kos,Karpie:2017bzm,Radyushkin:2017lvu} and the ``lattice cross section'' approach~\cite{Ma:2014jla,Ma:2014jga,Ma:2017pxb} are also developed.

The one-loop calculations of matching coefficient  in the cutoff scheme show that, the quasi quark PDF  contains linear power ultraviolet (UV)
divergences~\cite{Xiong:2013bka}.  These  divergences arise  from   Wilson-line's
self-energy.   Some suggestions have been made to solve the
problem, e.g., the non-dipolar Wilson line~\cite{Li:2016amo,Li:2017udt},  a mass counter term of Wilson
line~\cite{Chen:2016fxx}, as well as the auxiliary field
formalism~\cite{Green:2017xeu}.  All these methods are based on the fact
that the linear power divergences  arise from  Wilson lines' self-energy diagram. The renormalizability
of quasi quark distribution has also been proved to all orders  recently~\cite{Ji:2017oey,Ishikawa:2017faj}. Based on the multiplicative renormalizability of quasi-PDF, one can perform  nonperturbative  renormalization of quasi-PDF in the regularization-invariant  momentum-subtraction (RI/MOM) scheme~\cite{Martinelli:1994ty} , in which the UV divergence can be removed to all orders by a renormalization factor determined on the lattice~\cite{Chen:2017mzz,Green:2017xeu,Lin:2017ani,Stewart:2017tvs}.

Among various PDFs, the gluon
distribution function is a key quantity especially for the  processes like the production of   Higgs particles. It is also crucial for understanding the spin structure of
hadron. However, compared to the quasi quark distribution, the quasi gluon distribution is  less explored. A first calculation of perturbative coefficient for quasi gluon distribution
is presented in Ref.~\cite{Wang:2017qyg}, in which the results are presented in both the   cut-off and dimensional regularization (DR)
schemes. It is shown that  linear divergences exist even in the diagrams without any Wilson line, and thus these linear divergences can not be absorbed into  the renormalization of Wilson line. Hence the  gluon quasi PDF is much more involved.
In this work we focus on   power divergences in quasi gluon
distribution. We will show that, in the formalism of auxiliary field, the linear
divergences can be renormalized  by considering the contribution from possible
operator mixing, as well as the mass
counter term of Wilson line.

The rest of this paper is organized as follows. In
Sec.~\ref{sec:def}, we introduce a slightly modified definition of
quasi gluon distribution. In Sec.~\ref{sec:1loop}, we present the
perturbative calculation of quasi gluon distribution at one-loop
level. In Sec.~\ref{sec:aux},
the renormalization of linear divergence will be analyzed with auxiliary
field method. Then we propose an improved quasi gluon distribution
and calculate the matching coefficient in Sec.~\ref{sec:match}.
Sec.~\ref{sec:summary} is the summary and outlook of this work. In the appendix, we give some detailed calculations of the real diagrams with no Wilson lines involved.

\section{Definition of quasi gluon distribution}{\label{sec:def}}

The light-cone (or conventional) gluon distribution function is
defined by a non-local matrix element of two light-cone separated
gluon strength tensor
\begin{eqnarray}
    f_{g/H}(x,\mu) = \int \frac{d\xi^-}{2\pi x P^+} e^{-i\xi^- xP^+}  \langle P|G^+_{~i}(\xi^- n_+)
  W(\xi^- n_+,0; L_{n_+})G^{i +}(0) |P\rangle.\label{eq:def:lc}
\end{eqnarray}
In  the light-cone
coordinate,  a four-vector $a$ is expressed
as $a=(a^+, a^-,\vec{a}_{\perp})$, with $a^+=(a^0+a^z)/\sqrt{2}$ and
$a^-=(a^0-a^z)/\sqrt{2}$. $n_+=(0, 1, 0, 0)$ is a light-cone vector.
Similarly, the quasi gluon distribution can be defined by non-local
matrix element which is equal-time separated
\cite{Ji:2013dva,Ma:2017pxb,Wang:2017qyg}
\begin{align}
    \t f_{g/H}(x,P^z) = \int \frac{dz}{2\pi x P^z} e^{i z x P^z}  \langle P|G^z_{~i}(z n_z)
  W(z n_z,0; L_{n_z})G^{i z}(0) |P\rangle. \label{eq:quasi-PDF-def-old}
\end{align}
In quasi PDF we still adopt the Cartesian coordinates, where $n^{\mu}_z=(0,0,0,1)$ is an unite vector along $z$ direction.
In the above definitions, $W(x,y;C)$ denotes a Wilson line along
contour $C$ with two endpoints $x$ and $y$, the gluon field is in
adjoint representation and $L_n$ is a straight line with direction
vector $n$. The Wilson line can be
parametrized as
\begin{align}
  W(z_1,z_2, L_{n})=P \exp\left[-ig\int^1_0 d\lambda ~\dot{x}(\lambda)\cdot A(x(\lambda))\right],
\end{align}
where $P$ denotes that the operator exponential is path ordered. Any point
on $L_{n}$ can be expressed as
$x(\lambda)=\bar \lambda z_2 n+\lambda z_1 n$, where
$\bar\lambda\equiv 1-\lambda$, $x(0)=z_2 n$ and $x(1)=z_1 n$. $\dot
x^{\mu}(\lambda)\equiv d x^{\mu}(\lambda)/d\lambda$ is the velocity
vector of the contour.

According to Refs.~\cite{Ji:2013dva,Ma:2017pxb,Wang:2017qyg}, $i$ sums over the transverse directions. In this paper, we adopt
a modified definition, where the Lorentz index of gluon field
strength tensor is summed over all the directions except ``3'', i.e.,
\begin{eqnarray}
    \t f_{g/H}(x,\mu) = \int \frac{d z}{2\pi x P^z} e^{i z x P^z}  \langle P|G^z_{~{\mu}}(z n_z)
  W(z n_z,0;L_{n_z})G^{\mu z}(0) |P\rangle \ , \label{eq:def}
\end{eqnarray}
for $\mu=0,1,2$.

To see that Eq.~\eqref{eq:def} is a proper definition of quasi gluon
distribution, we begin with the energy-momentum
tensor of gluon field
\begin{align}
  T^{\mu\nu}= G^{\mu\lambda} G_{\lambda}^{~\nu}-\frac14 g^{\mu\nu}
  G^{\alpha\beta}G_{\beta\alpha}.\label{eq:emtensor}
\end{align}
The averaged value of $T^{\mu\nu}$ between  a hadron state $|P\rangle$ can
be expressed as
\begin{align}
  \langle P|T^{\mu\nu}|P\rangle=a(\mu)\left(P^{\mu}P^{\nu}-\frac14 M^2
  g^{\mu\nu}\right),\label{eq:lo}
\end{align}
in which $M$ is the hadron mass.
This tensor is symmetric and traceless with a scalar function
$a(\mu)$ characterizing the magnitude.  If we pick up the ``$++$'' component,
i.e. $\mu=\nu=+$, we have
\begin{align}
  \langle P|T^{++}|P\rangle=a(\mu)(P^+)^2. \label{eq:mom1}
\end{align}
The term proportional to $M^2$ disappears because $g^{++}=0$. On the
other hand, if we take $\mu=\nu=z$, then
\begin{align}
  \langle P|T^{zz}|P\rangle= a(\mu)\left[(P^z)^2-\frac14
  g^{zz}M^2\right]=a(\mu)\left[(P^z)^2+\frac14
  M^2\right]=a(\mu)(P^z)^2\left(1+O(M^2/(P^z)^2)\right).\label{eq:mom2}
\end{align}
Eqs.~\eqref{eq:mom1} and \eqref{eq:mom2} are related to the first
moment of light-cone and quasi gluon distribution respectively,
\begin{align}
  \int dx ~x f_{g/H}(x,\mu)&=\frac{1}{(P^+)^2}\langle
  P|T^{++}|P\rangle=a(\mu),\\
  \int dx ~x \t f_{g/H}(x,P^z,\mu)&=\frac{1}{(P^z)^2}\langle
  P|T^{zz}+\frac14 g^{zz}G^{\alpha\beta}G_{\alpha\beta}|P\rangle+{ O}\left(\frac{M^2}{(P^z)^2}\right)=a(\mu)+ {O}\left(\frac{\Lambda^2_{\mathrm{QCD}}}{(P^z)^2}, \frac{M^2}{(P^z)^2}\right).
\end{align}
Under the $P^z\to \infty$ limit,  one can obtain  $\int dx~ x
f_{g/H}(x,\mu)=\int dx~ x \t f_{g/H}(x,P^z,\mu)$. This relation can
be generalized to high order moments, and consequently, to the
distribution functions $f$ and $\t f$ themselves. It indicates that
Eq.~\eqref{eq:def} is a reasonable definition of quasi gluon
distribution. As a  comparison, if $\lambda$ is only summed
over transverse directions in Eq.~\eqref{eq:emtensor}, the tensor
will not be Lorentz covariant, and  then the above
arguments will no longer hold.

Actually, one can easily validate that both definitions in Eqs.~\eqref{eq:quasi-PDF-def-old} and \eqref{eq:def} will give the same results for the matching at tree-level. In addition,
the non-local operator $G^z_{~{\mu}}(z n_z)  W(z n_z,0;L_{n_z})G^{\mu z}(0)$ can be decomposed into the $\pm$ components:
\begin{eqnarray}
G^z_{~{\mu}}(z n_z)  W(z n_z,0;L_{n_z})G^{\mu z}(0) = \frac{1}{2} [G^{+}_{~{\mu}}(z n_z) -G^{-}_{~{\mu}}(z n_z)] W(z n_z,0;L_{n_z})[G^{\mu +}(0) -G^{\mu -}(0) ].
\end{eqnarray}
In the large $P^z$ limit, the $-$ component is suppressed and thus the operator for quasi-PDF recovers the same Lorentz structure with the light-cone PDF. The factor $1/2$ arises from the conversion of Euclidean and light-cone coordinates.

\section{One-loop corrections to quasi gluon distribution}{\label{sec:1loop}}

Based on the definition in Eq.~\eqref{eq:def}, we  will re-calculate the one-loop correction to quasi gluon
distribution. To perform  the matching, one can   replace the hadron state with a single parton state. In this  paper we are interested in the gluon-in-gluon case. The hadron state $|P\rangle$ can be replaced by an on-shell gluon state $|g(P)\rangle$, where $P$  is the gluon momentum. To regulate the UV divergence, we introduce an  UV cut-off $\Lambda$ on the transverse momentum. This cut-off corresponds to $1/a$ in LQCD, with  $a$ being the lattice spacing. The collinear divergence is regulated  by   a small gluon mass  $m_g$.  We will perform the perturbative calculation  in Feynman gauge.

\begin{figure}[!htp]
\begin{center}
\includegraphics[scale=0.6]{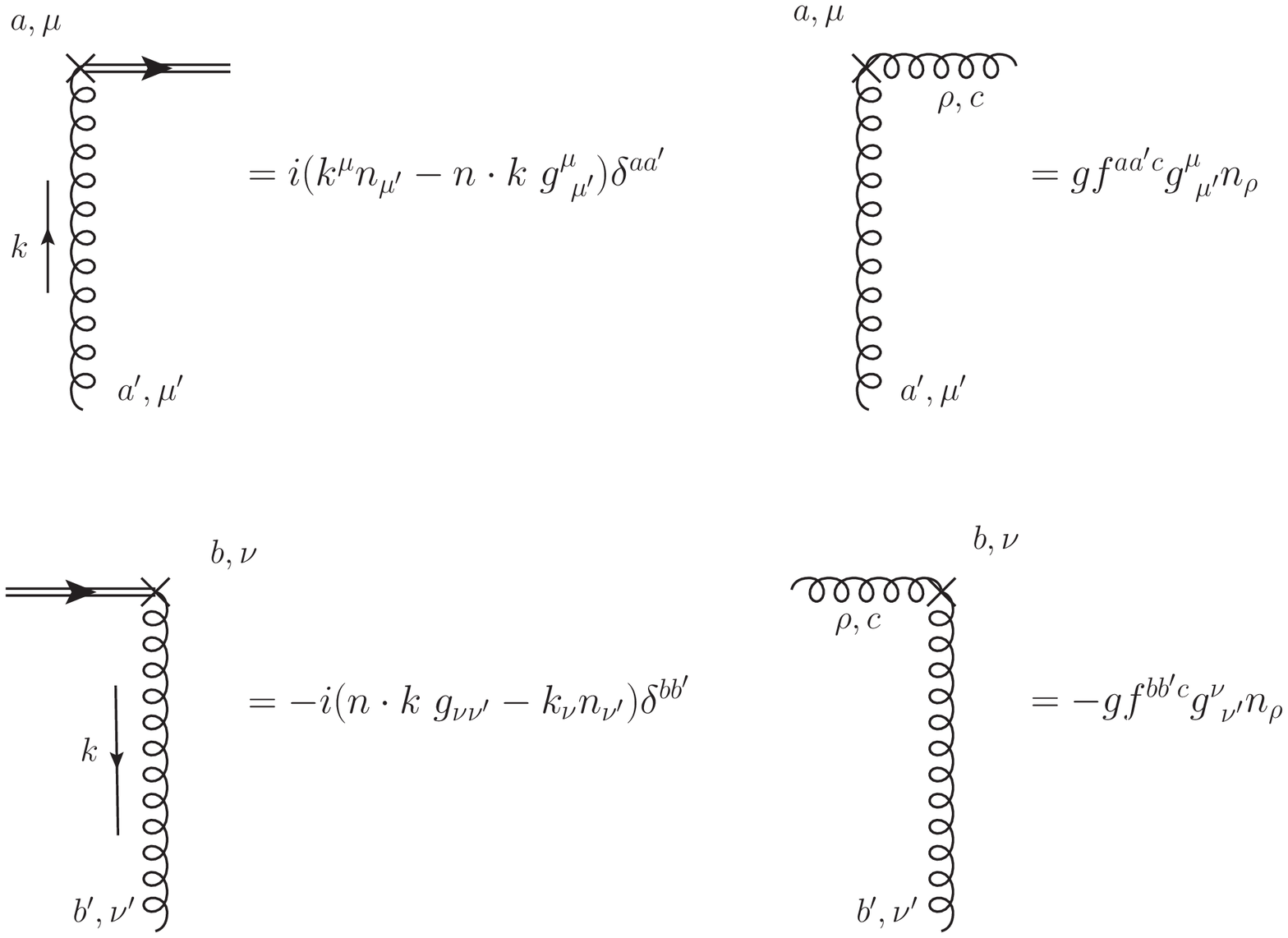}
\caption{The Feynman rules for non-local vertex from gluonium
operator. The left two sub-diagrams are from the Abelian part of the
gluon field strength tensor, while the right two are from the
non-Abelian part. } \label{fig:feynrules}
\end{center}
\end{figure}

The Feynman rules are listed in Fig.~\ref{fig:feynrules}. The left two sub-diagrams are from the Abelian part of the gluon field strength tensor, while the right two are from the non-Abelian part.  In these Feynman rules, we have explicitly separated the non-Abelian terms.

\begin{figure}[!htp]
\begin{center}
\includegraphics[scale=0.6]{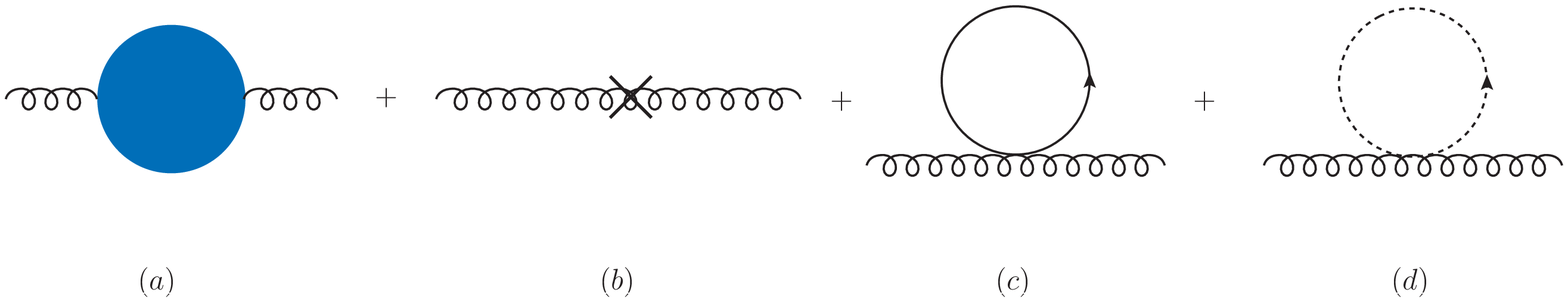}
\caption{One-loop self energy diagrams of gluon in lattice
perturbation theory. (a) denote the diagrams which have
corresponding diagrams in the  continuous limit. (b) is a counter-term
which arises from the measure of path integral. (c) and (d) are
tadpole diagrams involving a quark or ghost loop, which only exist
in lattice perturbation theory.} \label{fig:tadpoles}
\end{center}
\end{figure}

We firstly make some remarks on the self-energy corrections of the external
gluon. It is known for a long time that the UV cut-off scheme (on all the
components of internal momentum) leads to a quadratic divergence in
QCD vacuum polarization diagram, which breaks the Ward identity and
the gauge invariance.   An exception is the lattice perturbation
theory that provides  an UV cut-off scheme which preserve
the gauge invariance.  For a detailed review on lattice perturbation
theory, one can refer  to Ref.~\cite{Capitani:2002mp}.
The quadratic divergence in lattice perturbation
theory will  be canceled by  additional tadpole-like
diagrams, as well as the counter term which originates from the measure of
the path integral. These diagrams are shown in
Fig.~\ref{fig:tadpoles}, (a) has the  continuum analog and diverge like $1/a^2$. This $1/a^2$ divergence corresponds to the quadratic divergence in continuum
theory with a naive UV cut-off. The (b), (c) and (d) only exist in lattice theory but are essential to restore the gauge symmetry.   One can set the same renormalization scales (or the
UV cut-offs) for quasi and light-cone gluon distributions, then the
gluon self energy diagram contributes a same factor to light-cone
and quasi distributions, and will not contribute  to the matching coefficient. Based on the above discussions, it is not necessary to include the
gluon's self energy at present.

\begin{figure}[!htp]
\begin{center}
\includegraphics[scale=0.5]{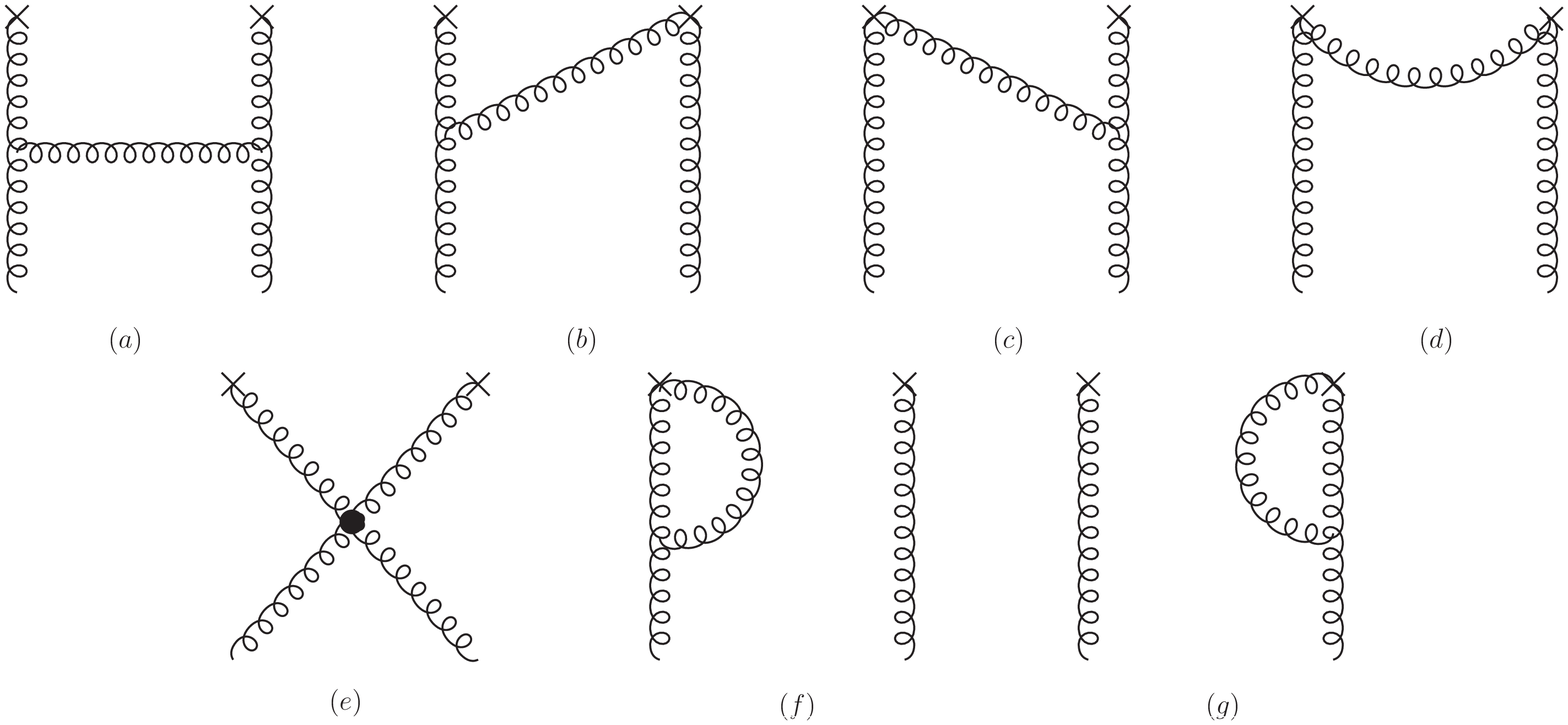}
\caption{One-loop corrections to quasi gluon distribution without
any Wilson line. The cross ``$\times$'' denotes the non-local vertex
from the operator structure of definition. } \label{fig:nowl}
\end{center}
\end{figure}

We start with  the calculation  of one-loop diagrams with no Wilson line. The Feynman diagrams   are shown in Fig.~\ref{fig:nowl}.
In Fig.~\ref{fig:nowl}(a), the non-local vertex is from the
 Abelian term of field strength tensor. A direct calculation gives
\begin{align}
 x\tilde f_{g/g}^{(1)}(x, P^z, \Lambda)\bigg\vert_{\mathrm{Fig}.\ref{fig:nowl}(a)}&=\frac{\alpha_s C_A}{2\pi }\left\{ \begin{aligned}
&\left(2 x^3-3 x^2+2 x-2\right) \ln \frac{x-1}{x}+2 x^2-\frac{5
x}{2}+3+\frac{\Lambda}{P^z}\ , & x>1
\\
&\left(2 x^3-3
x^2+2 x-2\right)\ln \frac{(1-x+x^2)m_g^2}{4x(1-x)(P^z)^2}\\
&~~+\frac{(2x-1)(4x^4-6x^3+10x^2-5x+2)}{2(1-x+x^2)}
+\frac{\Lambda}{P^z}, & 0<x<1\\
&-\left(2 x^3-3 x^2+2 x-2\right) \ln \frac{x-1}{x}-2 x^2+\frac{5
x}{2}-3+\frac{\Lambda}{P^z}\ .& x<0\end{aligned}
\right.\label{eq:quasi:cutoff}
\end{align}

Fig.~\ref{fig:nowl}(b, c) and (d) correspond to the contributions from
the non-Abelian term in the field strength tensor. Note that in
Ref.~\cite{Wang:2017qyg}, the non-Abelian term is combined with the
Wilson line. In the present work, the contributions
from these two parts are separated. This is more convenient to find   the source of
linear divergence. These diagrams give
\begin{align}
 &x\tilde f_{g/g}^{(1)}(x, P^z,\Lambda)\bigg\vert_{\mathrm{Fig}.\ref{fig:nowl}(b)}=\frac{\alpha_s C_A}{4\pi }\left\{
\begin{aligned}
&x(x+1)\ln\frac{x-1}{x}+2x-1-\frac{\Lambda}{P^z},~~~~&x>1\\
&x(x+1)\ln\frac{(1-x+x^2)m^2_g}{4x(1-x)(P^z)^2}+2x(x-1)+1-\frac{\Lambda}{P^z},~~~~&0<x<1\\
&-x(x+1)\ln\frac{x-1}{x}-2x+1-\frac{\Lambda}{P^z},~~~~&x<0
\end{aligned}\right.\\
&x\tilde f_{g/g}^{(1)}(x, P^z,\Lambda)\bigg\vert_{\mathrm{Fig}.\ref{fig:nowl}(c)}=x\tilde f_{g/g}^{(1)}(x, P^z,\Lambda)\bigg\vert_{\mathrm{Fig}.\ref{fig:nowl}(b)},
\end{align}
and
\begin{eqnarray}
 x\tilde f_{g/g}^{(1)}(x, P^z,\Lambda)\bigg\vert_{\mathrm{Fig}.\ref{fig:nowl}(d)}=\frac{\alpha_s C_A}{2\pi }\left\{ \begin{aligned}
&1-x+\frac{\Lambda}{P^z},~~~~&x>1\\
&x-1+\frac{\Lambda}{P^z},~~~~&0<x<1\\
&x-1+\frac{\Lambda}{P^z}.~~~~&x<0
\end{aligned}\right.
\end{eqnarray}

The four-gluon
coupling diagram,  depicted in Fig.~\ref{fig:nowl}(e), is
zero for light-cone distribution. However this diagram  contributes with a non-zero result
to the quasi distribution. This contribution is   linearly divergent:
\begin{eqnarray}
 x\tilde f_{g/g}^{(1)}(x, P^z,\Lambda)\bigg\vert_{\mathrm{Fig}.\ref{fig:nowl}(e)}=\frac{\alpha_s C_A}{2\pi }\left\{ \begin{aligned}
&\frac{x}{2}-\frac{\Lambda}{P^z},~~~~&x>1\\
&\frac{x}{2}-\frac{\Lambda}{P^z},~~~~&0<x<1\\
&-\frac{x}{2}-\frac{\Lambda}{P^z}.~~~~&x<0
\end{aligned}\right.
\end{eqnarray}

Fig.~\ref{fig:nowl}(f,g) are the virtual correction diagrams. Their contributions are proportional to the
tree-level result $\delta(1-x)$:
\begin{align}
 &x\tilde f_{g/g}^{(1)}(x, P^z,\Lambda)\bigg\vert_{\mathrm{Fig}.\ref{fig:nowl}(f)}=\delta(1-x)\frac{\alpha_s C_A}{4\pi }\int dy \left\{ \begin{aligned}
&(y+1)\ln\frac{y-1}{y},~~~~&y>1\\
&(y+1)\ln\frac{(1-y+y^2)m^2_g}{4y(1-y)(P^z)^2},~~~~&0<y<1\\
&-(y+1)\ln\frac{y-1}{y},~~~~&y<0
\end{aligned}\right.\\
&x\tilde f_{g/g}^{(1)}(x, P^z,\Lambda)\bigg\vert_{\mathrm{Fig}.\ref{fig:nowl}(g)}=x\tilde f_{g/g}^{(1)}(x, P^z,\Lambda)\bigg\vert_{\mathrm{Fig}.\ref{fig:nowl}(f)}.
\end{align}
Summing the contributions from Fig.~\ref{fig:nowl}(a)$\sim$(g)
together, we get the total contribution from Fig.~\ref{fig:nowl}:
\begin{align}
  x\tilde f_{g/g}^{(1)}(x,P^z,\Lambda)\bigg\vert_{\mathrm{Fig.}\ref{fig:nowl}}=&\frac{\alpha_s C_A}{2\pi }\left\{ \begin{aligned}
&(2x^3-2x^2+3x-2)\ln\frac{x-1}{x}+2x^2-2x+3,~~~~&x>1\\
&(2x^3-2x^2+3x-2)\ln\frac{(1-x+x^2)m^2_g}{4x(1-x)(P^z)^2}\nonumber\\
&~~+\frac{8 x^5-12 x^4+21 x^3-15 x^2+8 x-2}{2 \left(1-x+x^2\right)},~~~~&0<x<1\\
&-(2x^3-2x^2+3x-2)\ln\frac{x-1}{x}-2x^2+2x-3,~~~~&x<0
\end{aligned}\right.\nonumber\\
&+\delta(1-x)\frac{\alpha_s C_A}{2\pi }\int dy \left\{
\begin{aligned}
&(y+1)\ln\frac{y-1}{y},~~~~&y>1\\
&(y+1)\ln\frac{(1-y+y^2)m^2_g}{4y(1-y)(P^z)^2},~~~~&0<y<1\\
&-(y+1)\ln\frac{y-1}{y}.~~~~&y<0
\end{aligned}\right.
\end{align}
This result  is free of linear divergence, and  the linear divergences cancel with each other in Fig.~\ref{fig:nowl}(a)$\sim$(e). In the appendix, we give more details for this cancellation.

\begin{figure}[!htp]
\begin{center}
\includegraphics[scale=0.6]{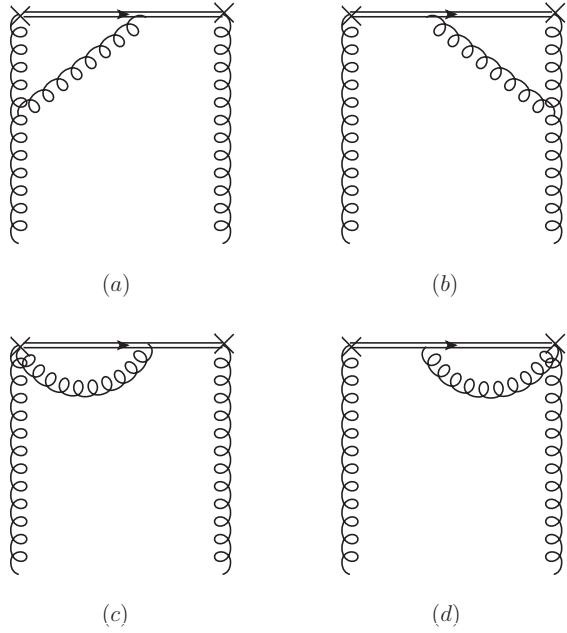}
\caption{One-loop corrections to quasi gluon distribution, which
involve one Wilson line and connects only one endpoint of internal gluon
line. The double line is the Wilson line. The cross ``$\times$''
denotes the non-local vertex from the operator structure of
definition.} \label{fig:wl1}
\end{center}
\end{figure}

\begin{figure}[!htp]
\begin{center}
\includegraphics[scale=0.6]{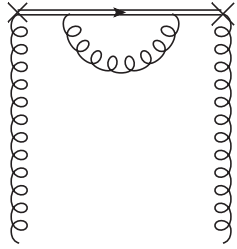}
\caption{One-loop correction to quasi gluon distributions, which
involves one Wilson line and both the endpoints of internal gluon line
connect to the Wilson line. The double line is the Wilson line.
The cross ``$\times$'' denotes the non-local vertex from the operator
structure of definition.} \label{fig:wl2}
\end{center}
\end{figure}

We come to the
diagrams which involve one Wilson line shown in
Figs.~\ref{fig:wl1} and \ref{fig:wl2}. In Fig.~\ref{fig:wl1}, only
one end of the internal gluon line is connected to the Wilson line,
while  both the ends of internal gluon line are
connected to the Wilson line in Fig.~\ref{fig:wl2}. A direct calculation shows that
\begin{align}
&x\tilde f_{g/g}^{(1)}(x, P^z,\Lambda)\bigg\vert_{\mathrm{Fig}.\ref{fig:wl1}(a)}=\frac{\alpha_s C_A}{4\pi }\left\{
\begin{aligned}
&\bigg[\frac{x(x+1)}{1-x}\ln\frac{x}{x-1}+\frac{1-2x}{1-x}+\frac{\Lambda}{(1-x)P^z}\bigg]_{\varoplus},&x>1\\
&\bigg[-\frac{x(x+1)}{1-x}\ln\frac{m^2_g(1-x+x^2)}{4x(1-x)(P^z)^2}-\frac{2x^2-2x+1}{1-x}+\frac{\Lambda}{(1-x)P^z}\bigg]_+,&0<x<1\\
&\bigg[-\frac{x(x+1)}{1-x}\ln\frac{x}{x-1}-\frac{1-2x}{1-x}+\frac{\Lambda}{(1-x)P^z}\bigg]_{\varominus},&x<0
\end{aligned}\right.\\
&x\tilde f_{g/g}^{(1)}(x, P^z,\Lambda)\bigg\vert_{\mathrm{Fig}.\ref{fig:wl1}(b)}=x\tilde f_{g/g}^{(1)}(x, P^z,\Lambda)\bigg\vert_{\mathrm{Fig}.\ref{fig:wl1}(a)},
\end{align}
and
\begin{eqnarray}
 x\tilde f_{g/g}^{(1)}(x, P^z,\Lambda)\bigg\vert_{\mathrm{Fig}.\ref{fig:wl1}(c)}=x\tilde f_{g/g}^{(1)}(x, P^z,\Lambda)\bigg\vert_{\mathrm{Fig}.\ref{fig:wl1}(d)}=\frac{\alpha_s C_A}{2\pi }\left\{
\begin{aligned}
&\bigg[-1-\frac{\Lambda}{(1-x)P^z}\bigg]_{\varoplus},~~~~&x>1\\
&\bigg[1-\frac{\Lambda}{(1-x)P^z}\bigg]_+,~~~~&0<x<1\\
&\bigg[1-\frac{\Lambda}{(1-x)P^z}\bigg]_{\varominus}.~~~~&x<0
\end{aligned}\right.
\end{eqnarray}
In the above expressions, three generalized functions are introduced to regulate the singular integral~\cite{Stewart:2017tvs}:
\begin{align}
  \int^{\infty}_1 dx~ [f(x)]_{\varoplus} T(x)&=\int^{\infty}_1 dx ~f(x)\left[T(x)-T(1)\right],\\
  \int^{1}_0 dx~ [f(x)]_{+} T(x)&=\int^{1}_0 dx ~f(x)\left[T(x)-T(1)\right],\\
   \int^{0}_{-\infty} dx~ [f(x)]_{\varominus} T(x)&=\int^0_{-\infty} dx ~f(x)\left[T(x)-T(1)\right].
\end{align}
Here  $T(x)$ is an arbitrary smooth   function. Then,
the total contribution from Fig.~\ref{fig:wl1} is given as
\begin{eqnarray}
 x\tilde f_{g/g}^{(1)}(x, P^z,\Lambda)\bigg\vert_{\mathrm{Fig}.\ref{fig:wl1}}=\frac{\alpha_s C_A}{2\pi }\left\{
\begin{aligned}
&\bigg[\frac{x(1+x)}{1-x}\ln\frac{x}{x-1}-\frac{1}{1-x}-\frac{\Lambda}{(1-x)P^z}\bigg]_{\varoplus},~~~~&x>1\\
&\bigg[-\frac{x(1+x)}{1-x}\ln\frac{(1-x+x^2)m^2_g}{4x(1-x)(P^z)^2}+\frac{1-2x^2}{1-x}-\frac{\Lambda}{(1-x)P^z}\bigg]_+,~~~~&0<x<1\\
&\bigg[-\frac{x(1+x)}{1-x}\ln\frac{x}{x-1}+\frac{1}{1-x}-\frac{\Lambda}{(1-x)P^z}\bigg]_{\varominus}.~~~~&x<0
\end{aligned}\right.\label{eq:linear:1wl}
\end{eqnarray}
One can find that the linear divergences  do not vanish.

At last we calculate the one-loop correction to the self energy of
Wilson line, which is presented by Fig.~\ref{fig:wl2}. The result is
\begin{eqnarray}
 x\tilde f_{g/g}^{(1)}(x, P^z,\Lambda)\bigg\vert_{\mathrm{Fig}.\ref{fig:wl2}}=\frac{\alpha_s C_A}{2\pi }\left\{
\begin{aligned}
&\bigg[\frac{1}{1-x}+\frac{\Lambda}{(1-x)^2P^z}\bigg]_{\varoplus},~~~~&x>1\\
&\bigg[\frac{1}{x-1}+\frac{\Lambda}{(1-x)^2P^z}\bigg]_+,~~~~&0<x<1\\
&\bigg[\frac{1}{x-1}+\frac{\Lambda}{(1-x)^2P^z}\bigg]_{\varominus}.~~~~&x<0
\end{aligned}\right.\label{eq:linear:wlse}
\end{eqnarray}
The linear divergence in Eq.~\eqref{eq:linear:wlse} has the similar
structure with the linear divergence in quasi quark distribution,
thus may  be renormalized in the same approach.

At last we note that there are also contributions from crossed  diagrams, which have  been discussed in Ref.~\cite{Wang:2017qyg}.
These contributions can be easily derived by using the property $\t
f(-x)=-\t f(x)$.

\section{Auxiliary field method and renormalization of linear
divergences}\label{sec:aux}

The light-cone and quasi parton distribution functions are defined
by the gauge invariant quark bi-linear or gluonium operator. A
Wilson line,  a path ordered exponential of a line integral
over gluon field along a contour, is essential to retaining the
gauge invariance. It has been known for a long time that loop
corrections for Wilson line lead to power divergences
\cite{Polyakov:1980ca,Dotsenko:1979wb,Brandt:1981kf}. The power
divergences can be summed to an exponential factor $\exp(-\pi/a
L(c))$ where $1/a$ is the UV cut-off and $L(c)$ is the length of
Wilson line. Based on this, the linear divergence in quasi quark
distribution are
cured~\cite{Chen:2016fxx,Ishikawa:2016znu,Green:2017xeu}, because
all the power divergences in quasi quark distribution are from the
Wilson line's self energy. In gluon quasi PDF, it has been shown in
Sec.~\ref{sec:1loop} that Wilson line's self energy is not the only
source of linear divergence. Since the Wilson line is a functional
of gluon field, the gluon operator and Wilson line might be
entangled in loop corrections.

 It is convenient to discuss gauge invariant non-local operator with the method of auxiliary
field~\cite{Gervais:1979fv}. In this method, the non-local operator
can be converted into local operators by introducing an auxiliary
field. The action of the theory is given by
\begin{align}
  S=S_{\mathrm{QCD}}+i\int^{\lambda}_0 \bar {\mathcal{Z}}\left(\frac{d}{d\lambda'}+i g \dot{x}^{\mu} A^a_{\mu} t^a
  \right)\mathcal{Z}
  d\lambda'+ i\bar{ \mathcal{Z}}(0) \mathcal{Z}(0),\label{eq:action}
\end{align}
where $S_{\mathrm{QCD}}$ is the action of QCD. Obviously the Wilson
line can be expressed by the two point Green's function of
$\mathcal{Z}-$ field
\cite{Samuel:1978iy,Gervais:1979fv,Arefeva:1980zd,Dorn:1986dt}
\begin{align}
  W(z_1,z_2; C)=\langle \mathcal{Z}(\lambda_1)\bar
  {\mathcal{Z}}(\lambda_2)\rangle_{\mathcal{Z}},\label{eq:2pointgreen}
\end{align}
$\langle\cdots\rangle_{\mathcal{Z}}$ denotes the expectation value respect to
$\mathcal{Z}$ only. Therefore, when $\mathcal{Z}$- field is
integrated out in Eq.~\eqref{eq:action}, all the internal
$\mathcal{Z}$ line is just the Wilson line.

With the $\mathcal{Z}$ field technique, the gauge invariant
non-local quark bi-linear operator is then given by
\cite{Samuel:1978iy,Gervais:1979fv,Arefeva:1980zd,Dorn:1986dt}
\begin{align}
  \bar\psi_i(z_2) W_{ij}(z_2,z_1; C)\psi_j(z_1)=\langle (\bar
  \psi(z_1)\mathcal{Z}(\lambda_1))(\bar{\mathcal{Z}}(\lambda_2)\psi(z_2))\rangle_{\mathcal{Z}},
\end{align}
and the gauge invariant non-local gluonium operator is
\begin{align}
  G^{a}_{\mu\nu}(z_1)W_{ab}(z_1,z_2; C)G^{b}_{\rho\sigma}(z_2)=\langle
  \left(G^{a}_{\mu\nu}(z_1)\mathcal{Z}_a(\lambda_1)\right)\left(\bar
  {\mathcal{Z}}_b(\lambda_2)G^{b}_{\rho\sigma}(z_2)\right)\rangle_{\mathcal{Z}}.
\end{align}
Now the problem of renormalizing the non-local operator has been
converted into the problem of renormalizing the local operators. The
operator is divided into two local operators,  which can be considered individually.

The operator $\Omega^{(1)}_{\mu\nu}=G^{a}_{\mu\nu}\mathcal{Z}_{a}$, which is a gauge invariant field strength tensor,
may mix with other operators under renormalization. The allowed mixing
operators are constrained by the gauge symmetry, or more strictly,
the BRST symmetry, as well as the dimensional constraint. It has been demonstrated  that  the $\Omega^{(1)}_{\mu\nu}$ can
only  mix with two operators under renormalization
\cite{Dorn:1981wa,Dorn:1986dt}
\begin{align}
  \Omega^{(1)}_{\mu\nu}&=G^{a}_{\mu\nu}\mathcal{Z}_{a},\\
\Omega^{(2)}_{\mu\nu}&=\Omega^{(1)}_{\mu\alpha}\frac{\dot{x}_{\alpha}\dot{x}_{\nu}}{\dot{x}^2}-\Omega^{(1)}_{\nu\alpha}\frac{\dot{x}_{\alpha}\dot{x}_{\mu}}{\dot{x}^2},\\
\Omega^{(3)}_{\mu\nu}&=|\dot{x}|^{-2}(\dot
{x}_{\mu}A^a_{\nu}-\dot{x}_{\nu}A_{\mu}^a)(D \mathcal{Z})_a,
\end{align}
with
\begin{align}
  (D \mathcal{Z})_a=\partial_{\lambda} \mathcal{Z}_a- g f_{abc} ~\dot{x}\cdot A_b \mathcal{Z}_c.
\end{align}
The Hermitian adjoint of these operators are
\begin{align}
  \bar{\Omega}^{(1)}_{\mu\nu}&=\bar{\mathcal{Z}}_{a}G^{a}_{\mu\nu},\\
\bar{\Omega}^{(2)}_{\mu\nu}&=\bar{\Omega}^{(1)}_{\mu\alpha}\frac{\dot{x}_{\alpha}\dot{x}_{\nu}}{\dot{x}^2}-\bar{\Omega}^{(1)}_{\nu\alpha}\frac{\dot{x}_{\alpha}\dot{x}_{\mu}}{\dot{x}^2},\\
\bar{\Omega}^{(3)}_{\mu\nu}&=(-\bar{\mathcal{Z}}\overleftarrow{D})_a|\dot{x}|^{-2}(\dot
{x}_{\mu}A^a_{\nu}-\dot{x}_{\nu}A_{\mu}^a),
\end{align}
with
\begin{align}
  (\bar{\mathcal{Z}}\overleftarrow{D})_a=-\partial_{\lambda} \bar{\mathcal{Z}}_a- g f_{abc} ~\dot{x}\cdot A_b \bar{\mathcal{Z}}_c.
\end{align}

After renormalization, the operator $\Omega^{(1)}_{\mu\nu}$ can be replaced by the operators $\Omega^{(i)(r.)}_{\mu\nu}(i=1,2,3)$ as
\begin{align}
\Omega^{(1)}_{\mu\nu}\rightarrow c_1 \Omega^{(1)(r.)}_{\mu\nu}+ c_2 \Omega^{(2)(r.)}_{\mu\nu}+ c_3\Omega^{(3)(r.)}_{\mu\nu},
\end{align}
where the $c_i~(i=1,2,3)$ are the mixing parameters, and the superscript $(r.)$ denotes that the $\mathcal{Z}$ field in the operators are renormalized.

However, the operator we are interested in is not the gluonium operator with full Lorentz indices. In the definition of quasi gluon
distribution, the operator should contract with $n_z$, which is the
unit tangent vector of the Wilson line contour. One can easily find that
$\Omega^{(1)}_{z\mu}=\Omega^{(2)}_{z\mu}$. Therefore, in the present work, $\Omega^{(1)}_{\mu\nu}$ and $\Omega^{(2)}_{\mu\nu}$ are not distinguishable.

\begin{figure}[!htp]
\begin{center}
\includegraphics[scale=0.6]{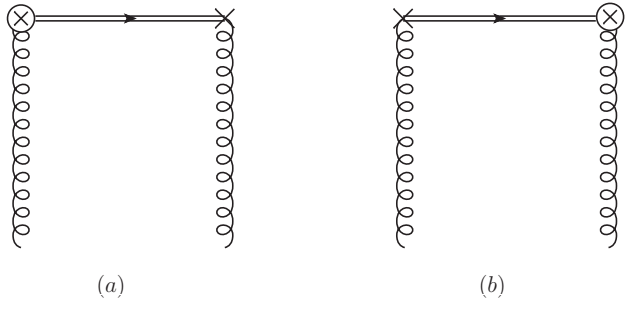}
\caption{The tree-level counter-term contribution, where
$\bigotimes$ denotes the counter-term from matrix elements of
$\Omega^{(3)(r.)}_{z\mu}$ or $\bar{\Omega}^{(3)(r.)}_{z\mu}$.}
\label{fig:cterm1}
\end{center}
\end{figure}

To study the renormalization of operator we should calculate the
tree level matrix element from the operator $\Omega^{(3)}_{\mu\nu}$.
Note that the operator we are interested in is
$\Omega^{(3)}_{z\mu}$. Since an arbitrary point on $W(z, 0, L_{n_z})$ can be expressed as $x(\lambda)=\lambda z n_z$, we have
\begin{align}
  \Omega^{(3)}_{z\mu}&=|\dot{x}|^{-2}(\dot{x}_z
  A^{a}_{\mu}-\dot{x}_{\mu}A^a_{z})(D\mathcal{Z})_a=-\frac{z}{|z|^2} A^a_{\mu}(D \mathcal{Z})_a.
\end{align}
Because the path in the Wilson line is along the $z$ direction, the $\mu$ component $(\mu=0, 1, 2)$
of $\dot{x}$ is zero.

Before calculating the tree-level matrix elements we make some
remarks on the renormalization of $\mathcal{Z}$ field. Under
renormalization, the $\mathcal{Z}$ field gets a renormalized mass
$\delta m_z$. Therefore, after renormalization, the equation of
motion for $\mathcal{Z}$ and $\bar{\mathcal{Z}}$ should be
\begin{align}
 \partial_{\lambda} \mathcal{Z}_a- g f_{abc} ~\dot{x}\cdot A_b \mathcal{Z}_c+\delta m_z |z|\mathcal{Z}^a=&0,\\
 \partial_{\lambda} \bar{\mathcal{Z}}_a+ g f_{abc} ~\dot{x}\cdot A_b \bar{\mathcal{Z}}_c-\delta m_z |z|\bar{\mathcal{Z}}^a=&0.\label{eq:eom:Z}
\end{align}
Then, the contraction of $\Omega^{(3)(r.)}_{z\mu}$ and external gluon state gives
\begin{align}
  &\langle \begC1{g}\conC{(p)|}\endC1{\Omega}^{(3)(r.)}_{z\mu}(z')=-\epsilon^*_{\mu}(p)e^{i p \cdot z'}\frac{z}{|z|^2}
 (D\mathcal{Z})_a=\delta m_z \frac{z}{|z|}\epsilon^*_{\mu}(p) e^{i p \cdot z'}\mathcal{
  Z}_a,\label{eq:contract}
\end{align}
where the equation of motion for $\mathcal{Z}$ (Eq.~\eqref{eq:eom:Z}) is used.

The contributions from $\Omega^{(3)(r.)}_{z\mu}$ at tree level are shown in Fig.~\ref{fig:cterm1}.
For Fig.~\ref{fig:cterm1}(a), with the contraction rule presented in Eq.~\eqref{eq:contract}, its contribution to quasi gluon distribution reads
\begin{align}
  x \t f_{g/g}(x,P^z,\Lambda)\bigg\vert_{\mathrm{Fig.}\ref{fig:cterm1}(a)}=&\int \frac{dz}{2\pi P^z} e^{i x P^z z}\langle g(p)|\langle(G^{z}_{~\mu} \mathcal{Z})(z n_z)
  \bar{\Omega}^{(3)(r.)\mu z}(0)\rangle_{\mathcal{Z}}|g(p)\rangle^{(0)}\nonumber\\
=&\frac{1}{2(N^2_c-1)}\delta^{a a'}\sum_{i}\epsilon_{\mu, i}(p)\epsilon^{*\mu}_i(p)\int \frac{dz}{2\pi
P^z} e^{i x
P^z z} i P^z e^{-i P^z z}(\delta m_z) \frac{-z}{|z|}\langle \mathcal{Z}_a(1)\bar{\mathcal{Z}}_{a'}(0)\rangle_{\mathcal{Z}} \nonumber\\
=&i P^z \delta m_z\int \frac{dz}{2\pi P^z} e^{i (x-1) P^z z}
\lim_{\epsilon\to
0}\frac{z}{|z|} e^{-\epsilon|z|}\nonumber\\
=&i P^z\delta m_z \int \frac{dz}{2\pi P^z} e^{i(x-1)P^z
z}\lim_{\epsilon\to 0}\int^{\infty}_0 d\alpha \frac{z e^{-\alpha
\epsilon^2-\frac{z^2}{4\alpha}}}{2\sqrt{\pi}\alpha^{\frac32}}\nonumber\\
=&i P^z\delta m_z \int \frac{dz}{2\pi P^z} e^{i(x-1)P^z
z}\lim_{\epsilon\to 0}\int^{\infty}_0
d\alpha\int^{+\infty}_{-\infty} d k^z \frac{k^z}{i\pi}
e^{-\alpha((k^z)^2+\epsilon^2)}(1-e^{-i k^z
z})\nonumber\\
=&P^z \delta m_z \int \frac{dz}{2\pi P^z} e^{i(x-1)P^z
z}\lim_{\epsilon\to 0}\int^{+\infty}_{-\infty}\frac{d k^z}{\pi}
\frac{k^z}{(k^z)^2+\epsilon^2}(1-e^{-i k^z z})\nonumber\\
=&\delta m_z \int^{+\infty}_{-\infty} \frac{d
k^z}{\pi}\frac{\delta((x-1)P^z)-\delta((x-1)P^z-k^z)}{k^z}\nonumber\\
=&\frac{\delta m_z}{\pi P^z}\left[\frac{1}{1-x}-\delta(1-x)\int
dy\frac{1}{1-y}\right]\nonumber\\
=&\frac{\delta m_z}{\pi
P^z}\left\{\begin{aligned}
&\left[\frac{1}{1-x}\right]_{\varoplus}, &x>1\\
&\left[\frac{1}{1-x}\right]_{+}, &0<x<1\\
&\left[\frac{1}{1-x}\right]_{\varominus}.  &x<0
\end{aligned}\right.\label{eq:counter1}
\end{align}
Here the superscript $(0)$ denotes the tree level matrix element.
In the above expression, $\langle \mathcal{Z}_a\bar{\mathcal{Z}}_{a'}\rangle_{\mathcal{Z}}$ is taken to be $\delta_{a a'}$ by recalling Eq.~\eqref{eq:2pointgreen} and also the fact that the matrix element is at tree level. It is interesting to notice that the divergence has the same structure   with Eq.~\eqref{eq:linear:1wl}.
Fig.~\ref{fig:cterm1}(b) gives the same contribution. Thus we have
\begin{align}
  x \t f_{g/g}(x, P^z, \Lambda)\bigg\vert_{\mathrm{Fig.}\ref{fig:cterm1}}=\frac{2\delta
  m_z}{\pi}\left\{\begin{aligned}&\left[\frac{1}{1-x}\right]_{\varoplus}, &x>1\\
  &\left[\frac{1}{1-x}\right]_{+}, &0<x<1\\
  &\left[\frac{1}{1-x}\right]_{\varominus}. &x<0\end{aligned}\right.
\end{align}


\begin{figure}[!htp]
\begin{center}
\includegraphics[scale=0.6]{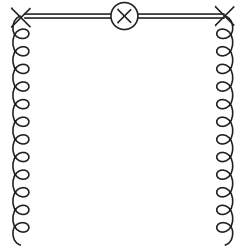}
\caption{The tree-level counter-term contribution, where
$\bigotimes$ denotes the mass counter-term of the Wilson line. }
\label{fig:cterm2}
\end{center}
\end{figure}


Now we calculate the contribution from Fig.~\ref{fig:cterm2}.
For a Wilson line in fundamental representation,
it has been shown
that~\cite{Polyakov:1980ca,Dotsenko:1979wb,Brandt:1981kf}
\begin{align}
  W(z_1,z_2; C)^{(r.)}=Z^{-1}\exp\left(-\delta m_z
  \|C(\lambda_1,\lambda_2)\|\right)W(z_1,z_2; C),
\end{align}
where $\|C\|$ denotes the length of the contour $C$. $Z$ is a renormalization constant.
Similarly, for a Wilson line $W(z,0,L_{n^z})$ in the adjoint representation, we have
\begin{align}
  W(z n_z,0; L_{n_z})^{(r.)}&=e^{-\delta m_z |z|}W(z n_z,0; L_{n_z})\nonumber\\
  &=1-\delta m_z |z|+W^{(1)}(z n_z,0; L_{n_z})+\cdots.
\end{align}
Therefore, with the same trick  in Ref.~\cite{Chen:2016fxx},
the leading order contribution from the counter term reads
\begin{align}
  &\int\frac{dz}{2\pi P^z}e^{i x P^z z}(-\delta m_z |z|)\langle
  g(p)|G^{z}_{~\mu}(z n_z)G^{\mu z}(0)|g(p)\rangle^{(0)}\nonumber\\
=&(-\delta m_z)\int\frac{dz}{2\pi P^z}e^{i x P^z z} |z| (P^z)^2 e^{-i
z P^z}\nonumber\\
=&(-\delta m_z)P^z\int \frac{dz}{2\pi} e^{i (x-1) P^z z}|z|\nonumber\\
=&-\delta m_z P^z\int\frac{d
k^z}{\pi}\frac{\delta((1-x)P^z)-\delta(k^z-(1-x)P^z)}{(k^z)^2}\nonumber\\
=&-\frac{\delta m_z}{\pi P^z}\left[-\frac{1}{(1-x)^2}+\delta(1-x)\int
dy\frac{1}{(1-y)^2}\right]\nonumber\\
=&\frac{\delta m_z}{\pi
P^z}\left\{\begin{aligned}&\left[\frac{1}{(1-x)^2}\right]_{\varoplus}, &x>1\\
&\left[\frac{1}{(1-x)^2}\right]_{+}, &0<x<1\\
&\left[\frac{1}{(1-x)^2}\right]_{\varominus}. &x<0\end{aligned}\right.\label{eq:counter2}
\end{align}

Now we turn back to the one-loop corrections, shown by Figs.~\ref{fig:wl1} and \ref{fig:wl2}. Fig.
\ref{fig:wl2} denotes the self-energy of Wilson line, and  the linear divergence in this diagram
 should be canceled by the one in Fig.~\ref{fig:cterm2}. From
Eqs.~\eqref{eq:linear:wlse} and \eqref{eq:counter2}, one can
determine that
\begin{align}
  \delta m_z=-\frac{\alpha_s C_A}{2\pi}(\pi \Lambda).\label{eq:deltam}
\end{align}

Fig.~\ref{fig:wl1} is the one-loop correction to the gluon-Wilson line vertex.
In the language of $\mathcal{Z}$ field formalism, it can also been
interpreted as the one-loop corrections to $\Omega^{(1)}_{z\mu}$.
The linear divergence in Fig.~\ref{fig:wl1} should be canceled by the one in
Fig.~\ref{fig:cterm1}. From Eqs.~\eqref{eq:linear:1wl} and
\eqref{eq:counter1} one can find  that the structures of linear
divergences are identically  the same. It indicates that all these linear
divergences can be renormalized by operator $\Omega^{(3)(r.)}_{z\mu}$,
with only the mixing parameter $c_3$ undetermined.  From
Eq.~\eqref{eq:deltam} and Eq.~\eqref{eq:linear:1wl}, we have
$c_3=-1/2$. Then, at one-loop level, all the linear divergences in quasi gluon distribution can be absorbed by $\delta m_z$, and
the matrix elements of  $\Omega^{(3)(r.)}_{z\mu}$.

The above calculation has been performed in the ``naive'' cut-off scheme, which may break  the gauge symmetry at first sight.
A complete analysis relies on the lattice perturbation theory. But  it is viable  to examine the gauge invariance from another viewpoint. In Ref.~\cite{Wang:2017qyg}, we have calculated
the matching coefficient in the dimensional regularization scheme. It is interesting to notice that
the differences between the cut-off and DR schemes   only reside in the linearly divergent terms, while the finite terms are the same. It is widely believed that gauge invariance is preserved in DR. In the present work, the linearly divergent terms have been absorbed into the matrix elements of the gauge invariant operators, leaving the finite terms consistent with the DR.
Therefore, it is   likely that   our results are  gauge invariant.

To further validate the above proposal, one needs  to investigate the cancelation of linear divergences for all loops. A complete proof is left for a future study, and here we outline the proof at
two-loop level.
Higher order corrections can be
generated by adding internal lines step by step from lower order
diagrams. The
contribution from a $n$-loop diagram with $k$ propagators can be generally written as
\begin{align}
  &\int d^4 l_1 d^4 l_2\cdots d^4 l_{n}\frac{N_1 N_2 \cdots N_j}{D_1 D_2\cdots
  D_k}\delta\left(\sum_{i=1}^n c_i
  l^z_{i}-(1-x)P^z\right)\nonumber\\
=&\int d l^z_{1} d l^z_{2} \cdots d l^z_{n}\delta\left(\sum_{i=1}^n c_i
  l^z_{i}-(1-x)P^z\right) \left[\int d^3 l_1 d^3 l_2\cdots d^3 l_n \frac{N_1 N_2
\cdots N_j}{D_1 D_2\cdots
  D_k}\right] ,
\end{align}
with $d^3 l_i\equiv d l^0_i d^2 l_{i\perp}$.
Here $N_i$ and $D_i$
denote the numerator and denominator
respectively, and  $c_i$ are some constants. In the discussions of Sec.~\ref{sec:1loop} and Sec.~\ref{sec:aux}, the $z$ component of loop momentum is either constrained by the $\delta$ function, or left unintegrated.
It has been shown that the linear divergences come from the integration over the ``0'' and transverse components of loop momenta.
Therefore, we separate  the $z$-components of the loop momenta $l_i$ and leave them
unintegrated. Then the loop integrals in the bracket [...] are all
three-dimensional (3D) integrals.
We define a few quantities for a general diagram as follows:
\begin{itemize}
  \item $L$: The number of loops.
  \item $F$: The number of internal fermions.
  \item $G$: The number of internal gluons and ghosts.
  \item $V_{3g}$: The number of three-gluon vertices and gluon-ghost vertices.
  \item $V_G$: The number of vertices corresponding to the Abelian part of field strength tensor. In addition, no external line should connects to these vertices.
\end{itemize}
With these quantities, we introduce the superficial degree of divergence $D$ for the 3D integrals, with $D=3L-F-2G+V_{3g}+V_G$. The propagator of Wilson line only involves the $z$-component of momentum, hence will not affect the 3D superficial degree of divergence.

There are three generic ways to generate  two-loop diagrams from the
one-loop diagram.
\begin{itemize}
\item{} None of the added internal line touches the Wilson line, which are shown in Fig.~\ref{fig:owl}.
This is the same with ordinary Feynman diagrams in QCD. For Fig.~\ref{fig:owl}(a, b), adding one gluon line will
introduce two three-gluon vertices and three gluon propagators, i.e., $G\to G+3$, $V_{3g}\to V_{3g}+2$ and $L\to L+1$. In this case $D$ will be decreased by 1. A linear divergence will reduce to a logarithm divergence, so there is no need to further subtract the linear divergence. Fig.~\ref{fig:owl}(c) contains gluon's self energy, although $D$ is increased by 1, the power divergence must be canceled by additional diagrams due to the gauge invariance, which has been discussed at the beginning of Sec.~\ref{sec:1loop}.

\begin{figure}[!htp]
\begin{center}
\includegraphics[scale=0.6]{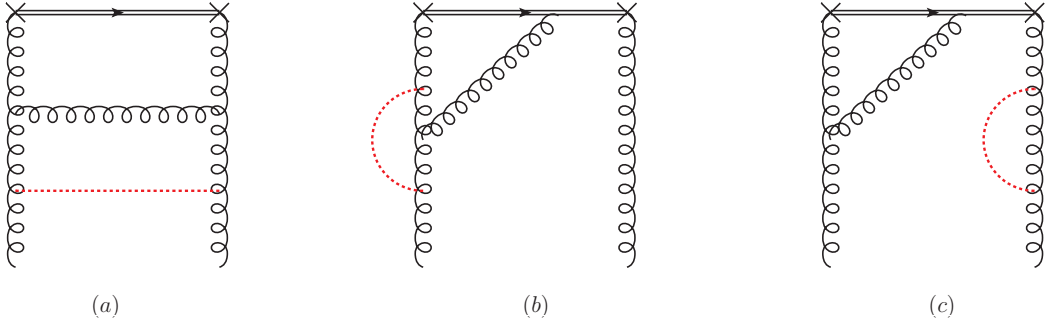}
\caption{Typical two-loops corrections to quasi gluon distribution,
in which the added internal line (denoted by dotted line) is not
connected to the Wilson line.} \label{fig:owl}
\end{center}
\end{figure}

\item{} Only one endpoint of the added gluon line is attached  to
Wilson line. This is shown in Fig.~\ref{fig:1wl}.  For Fig.~\ref{fig:1wl}(a), there will be two more
gluon propagators, one three-gluon vertex, and one Wilson line
propagator, i.e., $G\to G+2$, $V_{3g}\to V_{3g}+1$, and $L\to L+1$,
then $D$ will not be modified; For Fig.~\ref{fig:1wl}(b),
there will be two more
gluon propagators, one three-gluon vertex, and one Wilson line
propagator as well, but $V_G$ will be modified from 1 to 2,
then $D$ will be increased by 1. The power divergence in this diagram can be
viewed as the product of two linear divergences, which are from the
one-loop corrections of the gluon-Wilson line (or gluon-$\mathcal{Z}$) vertices.
In this case,
the power
divergence can be removed by adding contributions from
$\Omega^{(3)(r.)}_{\mu\nu}$. For Fig.~\ref{fig:1wl}(a) just one counter
term is needed, while for (b) we need two counter terms.


\begin{figure}[!htp]
\begin{center}
\includegraphics[scale=0.6]{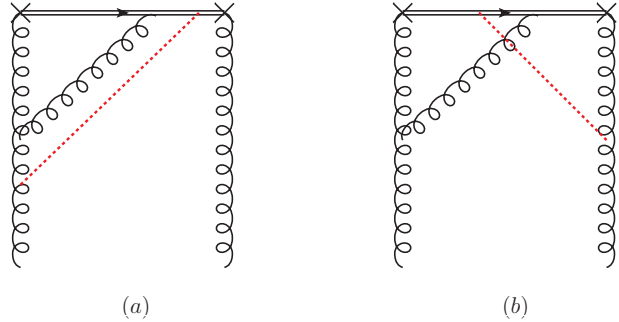}
\caption{Typical two-loops corrections to quasi gluon distribution,
in which only one endpoint of the added internal gluon line is connected
to the Wilson line.} \label{fig:1wl}
\end{center}
\end{figure}

\item{} Both the endpoints of the added gluon line touch the Wilson
line, which are shown in Fig.~\ref{fig:2wl}. In this case there will be one more
gluon propagator, i.e., $G\to G+1$, and two more Wilson line propagators, then $D$ will be increased by 1. This case can be explained as Wilson line's self
energy. The power divergence for this case should be cured by adding
mass counter term $\delta m_z$ of the Wilson line, as well as the corresponding counter term for
the linear divergence from the lower order diagram.

\begin{figure}[!htp]
\begin{center}
\includegraphics[scale=0.6]{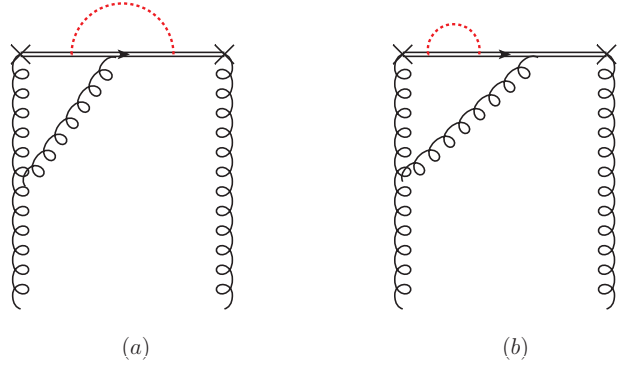}
\caption{Typical two-loops corrections to quasi gluon distribution,
in which both the endpoints of the added gluon line are connected to the
Wilson line.} \label{fig:2wl}
\end{center}
\end{figure}

\end{itemize}

The above discussions on two-loop corrections are conjectured to
hold to all orders, though we are lack of a complete all-order proof.

After discussing the linear divergence we add some remarks on the
logarithm divergence.
 From Sec.~\ref{sec:1loop}  one can find
that there is no logarithm divergence in the one-loop real
corrections.  By employing the auxiliary field approach the
renormalization of gauge invariant quark bi-linear and gluonium
operators has been discussed in DR scheme for a long
time~\cite{Craigie:1980qs,Dorn:1981wa,Dorn:1986dt}. Note that in DR
the linear divergence is not regularized. A remarkable
simplification is achieved when the gluonium operator is contracted
with the tangents of the contour, in this case, the operator will
not be renormalized \cite{Dorn:1981wa,Dorn:1986dt}, i.e.,
\begin{align}
  \dot {x}_{\mu} G^{\mu\nu\rho\sigma} \dot {x}_{\sigma}=\dot
  {x}_{\mu} (G^{\mu\nu\rho\sigma})^{(r.)}\dot {x}_{\sigma}
  \label{eq:ren:gluonium}
\end{align}
with
\begin{align}
  G^{\mu\nu\rho\sigma}=G^{\mu\nu}(z_1)W(z_1,z_2;C)G^{\rho\sigma}(z_2).
\end{align}
This  can be
traced back to the property of functional derivation of Wilson line.
The quasi gluon distribution belongs to this case.

It is also shown that the quasi distributions can   be defined
from other components of the correlators. For example, the quasi
quark distribution can   be defined by the matrix element of
$\bar \psi(z)\gamma^0\psi(0)$~\cite{Xiong:2013bka}. For gluon, one
can pick up any indices of ${\mu}$ and $\nu$ in the matrix elements
of $G^{\mu\rho}(z)W(z,0, L_{n_z})G_{\rho}^{~\nu}(0)$, in present work we choose
$\mu=\nu=z$. Eq.~\eqref{eq:ren:gluonium} indicates that, if one
defines quasi gluon distribution by specifying $\mu$ and $\nu$
components of gluonium operator, the renormalization of logarithm divergence
will be greatly simplified when the tangents of the contour are collinear to the
projection vector corresponding to these components; On the other
hand, if one modifies the contour of the Wilson line (e.g., in the
non-dipolar Wilson line approach~\cite{Li:2016amo}),   it is more convenient to modify the components of
gluon strength tensor correspondingly.

\section{Matching coefficient for improved gluon distribution}{\label{sec:match}}

With the discussions in Sec.~\ref{sec:aux}, we propose the subtracted quasi gluon distribution
as
\begin{align}
x \t f^{(r.)}_{g/H}(x,P^z,\Lambda)=\int\frac{d z}{2\pi P^z}e^{i x z P^z}\bigg\langle P\bigg|\bigg\langle\left(\Omega^{(1)(r.)z}_{~~~~~~~\mu}+c_3 \Omega^{(3)(r.)z}_{~~~~~~~\mu}\right)(z n_z)\left(\bar{\Omega}^{(1)(r.)\mu z}+c_3 \bar{\Omega}^{(3)(r.)\mu z}\right)(0)\bigg\rangle_{\mathcal{Z}}\bigg|P\bigg\rangle,
\end{align}
where $c_3=-1/2+O(\alpha_s)$ is the mixing parameter determined by perturbation theory. $\delta m_z$ and the matrix elements can be evaluated
non-perturbatively in LQCD. According to LaMET, one can expect a
factorization formula as
\begin{align}
  x\t f^{(r.)}_{g/H}(x, P^z ,\Lambda )=\int ^1_0 \frac{dy}{y}Z_{g i}\left(\frac{x}{y},
  \frac{\Lambda}{P^z}\right) y f_{i/H}(y, \Lambda).
\end{align}
In perturbation theory the $Z_{gg}$ can be expanded as
\begin{align}
  Z_{gg}\left(\xi, \frac{\Lambda}{P^z}\right)=\sum_{i=0}^{\infty}\left(\frac{\alpha_s}{2\pi}\right)^n
  Z^{(n)}_{gg}\left(\xi,\frac{\Lambda}{P^z}\right),
\end{align}
with $Z^{(0)}_{gg}(\xi)=\delta(1-\xi)$, $\xi\equiv x/y$.

The one-loop correction to light-cone gluon distribution has already
been calculated in Ref.~\cite{Wang:2017qyg}, with result
\begin{align}\label{eq:lightcone}
 x f_{g/g}^{(1)}(x, \Lambda)\bigg\vert_{\mathrm{total}.}=&\frac{\alpha_s
 C_A}{2\pi}\bigg\{\bigg[(2x^3-3x^2+2x-2)\ln\frac{(1-x+x^2)m^2_g}{\Lambda^2}-\frac{3x}{2(1-x+x^2)}+2x^2(x-1)+\frac72 x-2\bigg]\nonumber\\
 &-\frac{x^2(1+x)}{1-x}\ln\frac{(1-x+x^2)m^2_g}{\Lambda^2}-\delta(1-x)\int^1_0 dy
 \frac{1+y}{1-y}\ln\frac{\Lambda^2}{(1-y+y^2)m^2_g}\bigg\}\nonumber\\
 =&\frac{\alpha_s C_A}{2\pi }
\bigg\{(2x^3-2x^2+4x-1)\ln\frac{(1-x+x^2)
m^2_g}{\Lambda^2}+\bigg[\frac{x+1}{x-1}\ln\frac{(1-x+x^2)
m^2_g}{\Lambda^2}\bigg]_+\nonumber\\
&+2x^2(x-1)+\frac{7}{2}x-2-\frac{3x}{2(1-x+x^2)}\bigg\}
\end{align}
for $0<x<1$. The result is  0 for $x>1$ and $x<0$. On the other hand, the
one-loop correction to   gluon-in-gluon quasi distribution reads
\begin{eqnarray}\label{eq:quasi}
 x \t f_{g/g}^{(r.)(1)}(x, P^z,\Lambda)\bigg\vert_{total.}=\frac{\alpha_s C_A}{2\pi }\left\{
\begin{aligned}
&(2x^3-2x^2+4x-1)\ln\frac{x-1}{x}+\bigg[\frac{x+1}{x-1}\ln\frac{x-1}{x}\bigg]_{\varoplus}\\
&+2x^2-2x+3,~~~~&x>1\\
&(2x^3-2x^2+4x-1)\ln\frac{(1-x+x^2)m^2_g}{4x(1-x)(P^z)^2}\\
&+\bigg[\frac{x+1}{x-1}\ln\frac{(1-x+x^2)m^2_g}{4x(1-x)(P^z)^2}+\frac{2x^2}{x-1}\bigg]_+\\
&+\frac{8 x^5-12 x^4+21 x^3-15 x^2+8
x-2}{2 \left(1-x+x^2\right)},~~~~&0<x<1\\
&-(2x^3-2x^2+4x-1)\ln\frac{x-1}{x}-\bigg[\frac{x+1}{x-1}\ln\frac{x-1}{x}\bigg]_{\varominus}\\
&-2x^2+2x-3.~~~~&x<0
\end{aligned}\right.
\end{eqnarray}

From Eqs.~\eqref{eq:lightcone} and \eqref{eq:quasi}, one can determine the
matching coefficient at one-loop level,
\begin{align}
  Z^{(1)}_{gg}\bigg(\xi, \frac{\Lambda}{P^z}\bigg)=C_A\left\{
\begin{aligned}
&(2\xi^3-2\xi^2+4\xi-1)\ln\frac{\xi-1}{\xi}+\bigg[\frac{\xi+1}{\xi-1}\ln\frac{\xi-1}{\xi}\bigg]_{\varoplus}+2\xi^2-2\xi+3,~~~~&\xi>1\\
&(2\xi^3-2\xi^2+4\xi-1)\ln\frac{\Lambda^2}{4\xi(1-\xi)(P^z)^2}+\bigg[\frac{\xi+1}{\xi-1}\ln\frac{\Lambda^2}{4\xi(1-\xi)(P^z)^2}\bigg]_+\\
&+\bigg[\frac{2\xi^2}{\xi-1}\bigg]_++2\xi^3+\xi+1,~~~~&0<\xi<1\\
&-(2\xi^3-2\xi^2+4\xi-1)\ln\frac{\xi-1}{\xi}-\bigg[\frac{\xi+1}{\xi-1}\ln\frac{\xi-1}{\xi}\bigg]_{\varominus}-2\xi^2+2\xi-3.~~~~&\xi<0
\end{aligned}\right.
\end{align}
This coefficient has only the logarithmic dependent on the cut-off $\Lambda$. It is
free of IR divergence, which indicates that the LaMET factorization
holds for gluon distribution at one-loop level.

\section{summary and outlook}
\label{sec:summary}

In this work we have  discussed   power divergences  in the quasi gluon
distribution.  The gluon PDF is parametrized by two gluon field strength tensors contracted with each other.  From  the viewpoint of covariance, we proposed in the contraction to sum all components  of Lorentz indices instead of only transverse ones. With this definition, we have re-calculated  the gluon-in-gluon quasi distribution function at one-loop level, in which the Abelian and non-Abelian contributions are separated.

We found that power divergences cancel in diagrams without Wilson line,  and not all linear divergences  can be attributed into the renormalization of Wilson line.
Employing auxiliary field method,  we express the Wilson line as the expectation value of the product of two auxiliary fields. We found that a new operator can mix with the gauge invariant gluon field strength tensor, and  pointed out that the power divergence can be absorbed into the matrix elements of this newly introduced operator at one-loop level.  This cancelation of power divergence is conjectured to
hold to all orders.

We note that  since one can identify the $\mathcal{Z}$ field as a Wilson line $\mathcal{Z}(z)=W(z,\infty,L_{n_z})$,  this new operator is a product of  covariant derivatives of Wilson line.
We then have used an improved quasi gluon distribution, and determined the matching
coefficient  at   one-loop level. This matching  coefficient   is IR
finite and free of UV power divergence. Therefore, the present work provides the possibility of extracting gluon distributions from
LaMET and LQCD calculation.

We have to note that it is still a challenge of evaluating gluon PDF on the lattice and the present work leaves substantial room for improvement.
\begin{itemize}
\item In this work, we only consider the quasi gluon PDF in continuum QCD, the non-local gluon operators are constructed by the gluon field strength tensor and Wilson lines are under adjoint representation.  To simulate quasi gluon PDF on the lattice, one should define quasi PDF and perform the matching calculation in lattice gauge theory, in which the gauge field is expressed by the Wilson loops under fundamental representation. For a practical lattice evaluation, one needs to define the gauge invariant non-local gluon operators by the renormalized Wilson loops under fundamental representation. The mixing of the gauge invariant gluon operators should be considered as well.
\item Since the quasi PDF should be evaluated on the lattice while the light-cone PDF is defined in continuum, one has to match the light-cone PDF in continuum to the quasi PDF calculated with the lattice regularization. Furthermore, to arrive at a reliable result, renormalizing the matrix elements on the lattice is also a crucial step. Two approaches have been employed to renormalize the quasi quark PDF, one is the lattice perturbation theory~\cite{Chen:2016fxx,Carlson:2017gpk,Xiong:2017jtn,Constantinou:2017sej}, another one is the nonperturbative renormalization approach, e.g., the RI/MOM scheme~\cite{Alexandrou:2017huk,Chen:2017mzz,Lin:2017ani,Green:2017xeu,Stewart:2017tvs}. The renormalized matrix element in RI/MOM is independent of UV regularization, thus one can match the lattice evaluated quasi PDF to the $\overline{\mathrm{MS}}$ renormalized light-cone PDF, and work out the matching coefficient in DR, instead of the naive cut-off. However, applying RI/MOM to gluon operators is not straightforward for quasi gluon PDF, the reason is that the multiplicative renormalizability is not clear for non-local gluon operator, during to the operator mixing discussed in this paper. At the moment, a perturbative renormalization with lattice perturbation theory will be useful in practical extraction of gluon PDF.
\end{itemize}
These issues will be addressed in the forthcoming work.

\section*{Acknowledgments}

We  are grateful to  J. W. Chen, Y. Jia, J. P. Ma, and  R. L. Zhu for inspiring discussions, and T. Ishikawa, X. D. Ji,  H. N. Li, J. Xu, Y. B. Yang and J. H. Zhang
for valuable suggestions. W.W. thanks Cai-Dian L\"u and Qiang Zhao for their hospitality when this work is finalized.
This work is supported  in part   by National  Natural
Science Foundation of China under Grant
 No.11575110, 11655002, 11735010,  Natural  Science Foundation of Shanghai under Grant  No.~15DZ2272100 and No.~15ZR1423100, Shanghai Key Laboratory for Particle Physics and Cosmology, and  by  MOE  Key Laboratory for Particle Physics, Astrophysics and Cosmology.

\appendix

\section{Cancelation of the linear divergences in no-Wilson line diagrams}

In Sec.~\ref{sec:1loop}, we have concluded that the linear divergences cancel between the diagrams which with no Wilson line (Fig.~\ref{fig:nowl}). In the following discussion, we will show that how the cancelation works.

The UV divergence appears when loop momenta go to infinity. Since the $z$ component of loop momentum in one-loop real correction is constrained by the $\delta$ function, the UV divergence is from the region $k\sim (\Lambda, \Lambda, \Lambda, P^z)$, where $k$ is the loop momentum and $\Lambda\gg P^z$ is the largest scale. In this region, one can neglect $k^z$ as well as the external momentum $P$. Then, under this approximation, $k$ is replaced by $\t{k}=(k^0, k^x, k^y, 0)$. To compare with the original definition in Refs.~\cite{Ji:2013dva,Ma:2017pxb,Wang:2017qyg} and the new definition in Eq.~\eqref{eq:def}, we calculate the correlation matrix element
\begin{align}
  \mathcal{M}^{\mu\nu}=\int\frac{d z}{2\pi P^z}e^{i x z P^z}\langle g(P)|G^{z\mu}(z n_z)G^{\nu z}(0)|g(P)\rangle.
\end{align}
Since we are interested in the no-Wilson diagrams, the Wilson line has been eliminated.

In the region $k\sim (\Lambda, \Lambda, \Lambda, P^z)$, for Fig.~\ref{fig:nowl}, we have
\begin{align}
  \mathcal{M}^{\mu\nu}\bigg\vert_{\mathrm{Fig.}\ref{fig:nowl}(a)}&\sim -2 i g^2 C_A \frac{n^2}{P^z} \int\frac{d^3 \t{k}}{(2\pi)^4} \frac{\t{k}^{\mu}\t{k}^{\nu}k^2_{\perp}}{(\t{k}^2)^3}+O(\Lambda^0),\\
  \mathcal{M}^{\mu\nu}\bigg\vert_{\mathrm{Fig.}\ref{fig:nowl}(b)}&\sim  i g^2 C_A \frac{n^2}{P^z} \int\frac{d^3 \t{k}}{(2\pi)^4} \frac{k^{\mu}_{\perp}\t{k}^{\nu}}{(\t{k}^2)^2}+O(\Lambda^0),\\
  \mathcal{M}^{\mu\nu}\bigg\vert_{\mathrm{Fig.}\ref{fig:nowl}(c)}&\sim i g^2 C_A \frac{n^2}{P^z} \int\frac{d^3 \t{k}}{(2\pi)^4} \frac{\t{k}^{\mu}k^{\nu}_{\perp}}{(\t{k}^2)^2}+O(\Lambda^0),\\
  \mathcal{M}^{\mu\nu}\bigg\vert_{\mathrm{Fig.}\ref{fig:nowl}(d)}&\sim -i g^2 C_A \frac{n^2}{2P^z} \int\frac{d^3 \t{k}}{(2\pi)^4} \frac{g^{\mu\nu}_{\perp}}{\t{k}^2}+O(\Lambda^0),\\
  \mathcal{M}^{\mu\nu}\bigg\vert_{\mathrm{Fig.}\ref{fig:nowl}(e)}&\sim i g^2 C_A \frac{n^2}{P^z} \int\frac{d^3 \t{k}}{(2\pi)^4} \frac{\t{k}^{\mu}\t{k}^{\nu}}{(\t{k}^2)^2}+O(\Lambda^0).
\end{align}

By contracting $\mathcal{M}^{\mu\nu}$ with $g_{\mu\nu}$ we arrive at the definition of quasi gluon PDF adopted in this work. Then, from the above results, we have
\begin{align}
  x\t{f}(x, P^z, \Lambda) \bigg\vert_{\mathrm{Fig.}\ref{fig:nowl}(a)}&\sim -2 i g^2 C_A \frac{n^2}{P^z} \int\frac{d^3 \t{k}}{(2\pi)^4} \frac{k^2_{\perp}}{(\t{k}^2)^2}+O(\Lambda^0),\\
  x\t{f}(x, P^z, \Lambda)\bigg\vert_{\mathrm{Fig.}\ref{fig:nowl}(b)}&\sim  i g^2 C_A \frac{n^2}{P^z} \int\frac{d^3 \t{k}}{(2\pi)^4} \frac{k^2_{\perp}}{(\t{k}^2)^2}+O(\Lambda^0),\\
  x\t{f}(x, P^z, \Lambda)\bigg\vert_{\mathrm{Fig.}\ref{fig:nowl}(c)}&\sim i g^2 C_A \frac{n^2}{P^z} \int\frac{d^3 \t{k}}{(2\pi)^4} \frac{k^2_{\perp}}{(\t{k}^2)^2}+O(\Lambda^0),\\
  x\t{f}(x, P^z, \Lambda)\bigg\vert_{\mathrm{Fig.}\ref{fig:nowl}(d)}&\sim -i g^2 C_A \frac{n^2}{P^z} \int\frac{d^3 \t{k}}{(2\pi)^4} \frac{1}{\t{k}^2}+O(\Lambda^0),\\
  x\t{f}(x, P^z, \Lambda)\bigg\vert_{\mathrm{Fig.}\ref{fig:nowl}(e)}&\sim i g^2 C_A \frac{n^2}{P^z} \int\frac{d^3 \t{k}}{(2\pi)^4} \frac{1}{\t{k}^2}+O(\Lambda^0).
\end{align}
One can find that in the region $k\sim (\Lambda, \Lambda, \Lambda, P^z)$,
\begin{align}
   &x\t{f}(x, P^z, \Lambda) \bigg\vert_{\mathrm{Fig.}\ref{fig:nowl}(a)}+ x\t{f}(x, P^z, \Lambda) \bigg\vert_{\mathrm{Fig.}\ref{fig:nowl}(b)}+ x\t{f}(x, P^z, \Lambda) \bigg\vert_{\mathrm{Fig.}\ref{fig:nowl}(c)}\sim O(\Lambda^0),\\
   &x\t{f}(x, P^z, \Lambda) \bigg\vert_{\mathrm{Fig.}\ref{fig:nowl}(d)}+ x\t{f}(x, P^z, \Lambda) \bigg\vert_{\mathrm{Fig.}\ref{fig:nowl}(e)}\sim O(\Lambda^0).
\end{align}
Therefore, the linear divergences cancel between the no-Wilson line diagrams under the definition proposed in this work.

On the other hand, if $\mathcal{M}^{\mu\nu}$ is contracted with $g_{\perp \mu\nu}$, we return to the definition proposed by Refs.~\cite{Ji:2013dva,Ma:2017pxb,Wang:2017qyg}. In the region $k\sim(\Lambda, \Lambda, \Lambda, P^z)$, we have
\begin{align}
  g_{\perp \mu\nu}\mathcal{M}^{\mu\nu}\bigg\vert_{\mathrm{Fig.}\ref{fig:nowl}(a)}&\sim -2 i g^2 C_A \frac{n^2}{P^z} \int\frac{d^3 \t{k}}{(2\pi)^4} \frac{(k^2_{\perp})^2}{(\t{k}^2)^3}+O(\Lambda^0),\\
  g_{\perp \mu\nu}\mathcal{M}^{\mu\nu}\bigg\vert_{\mathrm{Fig.}\ref{fig:nowl}(b)}&\sim  i g^2 C_A \frac{n^2}{P^z} \int\frac{d^3 \t{k}}{(2\pi)^4} \frac{k^2_{\perp}}{(\t{k}^2)^2}+O(\Lambda^0),\\
  g_{\perp \mu\nu}\mathcal{M}^{\mu\nu}\bigg\vert_{\mathrm{Fig.}\ref{fig:nowl}(c)}&\sim i g^2 C_A \frac{n^2}{P^z} \int\frac{d^3 \t{k}}{(2\pi)^4} \frac{k^2_{\perp}}{(\t{k}^2)^2}+O(\Lambda^0),\\
  g_{\perp \mu\nu}\mathcal{M}^{\mu\nu}\bigg\vert_{\mathrm{Fig.}\ref{fig:nowl}(d)}&\sim -i g^2 C_A \frac{n^2}{P^z} \int\frac{d^3 \t{k}}{(2\pi)^4} \frac{1}{\t{k}^2}+O(\Lambda^0),\\
  g_{\perp \mu\nu}\mathcal{M}^{\mu\nu}\bigg\vert_{\mathrm{Fig.}\ref{fig:nowl}(e)}&\sim i g^2 C_A \frac{n^2}{P^z} \int\frac{d^3 \t{k}}{(2\pi)^4} \frac{k^2_{\perp}}{(\t{k}^2)^2}+O(\Lambda^0).
\end{align}
Totally,
\begin{align}
  g_{\perp \mu\nu}\mathcal{M}^{\mu\nu}\bigg\vert_{\mathrm{Fig.}\ref{fig:nowl}}\sim i g^2 C_A\frac{n^2}{P^z}\int\frac{d^3 \t{k}}{(2\pi)^4}\frac{k^2_0 k^2_{\perp}-k^2_0 k^2_0}{(\t{k}^2)^3}+O(\Lambda^0)\sim O(\Lambda^1),
\end{align}
which indicates that the linear divergence will not cancel between diagrams with no Wilson line under the original definition.

\end{document}